\documentclass[11pt,a4paper]{article}
\usepackage{jheppub}

\usepackage{epsfig,multicol,bbm}
\usepackage[sort&compress,numbers]{natbib}
\usepackage{graphicx}
\newcommand{\beq}{\begin{equation}}
\newcommand{\eeq}{\end{equation}}
\newcommand{\beqa}{\begin{eqnarray}}
\newcommand{\eeqa}{\end{eqnarray}}
\newcommand{\nn}{\nonumber}
\newcommand{\brem}{bremsstrahlung\ }

\title{A Forward Branching Phase Space Generator for Hadron Colliders}

\author[a]{Terrance M.~Figy }
\affiliation[a]{Department of Mathematics, Statistics, and Physics,
Wichita State University, Wichita, Kansas, USA}
\author[b]{and Walter T.~Giele}
\affiliation[b]{Theory Group, Fermilab, Batavia, USA}
\emailAdd{terrance.figy@wichita.edu}
\emailAdd{giele@fnal.gov}

\abstract{
In this paper we develop a projective phase space generator 
appropriate for hadron collider geometry.
The generator integrates
over \brem events which project back to a single, fixed
Born event. The projection is dictated
by the experimental jet algorithm allowing for the forward branching
phase space generator to integrate out the jet masses
and initial state radiation.  When integrating over the virtual and
\brem amplitudes this results in a single $K$-factor,
assigning an event probability to each Born event.
This $K$-factor is calculable as a perturbative expansion in the strong
coupling constant.

One can build observables from the Born kinematics,
giving identical results to traditional observables as long
as the observable does not depend on the infrared sensitive
jet mass or initial state radiation.
}

\preprint{preprint FERMILAB-PUB-18-187-T}
\arxivnumber{1806.09678}
\notoc
\begin{document}
\maketitle

\section{Introduction}

Great advances have been made over the last decades in calculating scattering amplitudes
contributing to higher order corrections for processes including
jets
In contrast the Monte Carlo methods of integrating over phase space
have not evolved much, relying instead on
advances in computer power. Given this, there was no
urgency to develop improved phase space integration methods.
However, the required precision of predictions needed for
current and future experimental analyses makes the development of improved phase space integration
methods a crucial task.  In this paper, we will investigate the promising avenue of projective phase space 
generators~\cite{Giele:2011tm,Campbell:2012cz,Cacciari:2015jma,Giele:2015sva,Currie:2018fgr,Martini:2015fsa,Martini:2017ydu,Martini:2018imv}. 

The basics of Monte Carlo integration are simple.
Take as an example the process $pp\rightarrow V + 1$ jet at
Next-to-Leading (NLO) order.
One generates random phase space events for the $V +1$ massless
particle production, assisted by importance sampling as implemented
in, for example, VEGAS~\cite{Lepage:1977sw}, and evaluates the Born and dressed Virtual amplitudes to
obtain the virtual weight. The observable under study is calculated
from the momenta and the weight is binned accordingly in a histogram.
This procedure is repeated for the $V+2$ massless particle production.
The events are randomly generated, used to evaluate the regulated
\brem amplitude, a jet algorithm is applied and the observable
calculated. Finally, the \brem amplitude weight is added to the corresponding
bin in the histogram.

In this procedure the use of finite width bins is essential to ensure
the cancellations between the virtual and \brem contributions, needed to
obtain a physical prediction. There are many drawbacks to this method.
First of all, the bin edges can cause issues in the cancellations leading
to instabilities which might need to be regulated.
Secondly, the higher order phase space is extended with respect to the Born phase space.
Consequently higher order corrections for an observable cover kinematic regions not
accessible to the Born results, leading to instabilities in the predictions.
Finally, in contrast to Leading Order (LO), one cannot define a weight at higher orders
to a single event ~\cite{Chung:2012rq,Goetz:2012uz,Eynck:2001en,Catani:1996vz}.
At LO each $V +$ 1 jet event has a probabilistic weight associated with it, allowing the
events to be unweighted. This is an important step as the subsequent
showering and detector simulations are potentially time expensive.
At NLO and beyond this is no longer possible,
making higher order predictions far more cumbersome and time consuming~\cite{Frixione:2002ik,Frixione:2007vw}.

To alleviate these issues and obtain more efficient Monte Carlo integration methods
for higher order calculations one has to change the basic method of phase space
integration.
A single $pp\rightarrow V +$ jet(s) event is a well defined
multi-dimensional observable with respect to the reconstructed final state object.
A scattering probability should be calculable order-by-order in the
perturbative expansion. This single observed event is semi-inclusive
with respect to the underlying partonic
event due to the use of a jet algorithm necessary to obtain the jet final state.
We have to integrate the partonic
\brem events over the jet cones and over the initial state radiation not associated with jets
under the constraint of a fixed $V +$ jet(s) event. In doing this one will
obtain an infra-red stable, physical prediction order-by-order in perturbation theory.
Such a method would allow for the generation of unweighted higher order events, possess no phase space extensions and
possess no other sources of instabilities in the predictions. In other words, all the artificial
instabilities introduced by the integration method and observable definitions are removed.
In two previous papers the methodology of Forward Branching Phase (FBPS) generators was
developed to some extend. The first paper~\cite{Giele:2011tm} 
develops the principles of the method from a more theoretical perspective. It uses 
a a $3\rightarrow 2$ jet clustering algorithm instead of a $2\rightarrow 1$ clustering 
algorithm used in experiments. While from a theoretical point of view the 
$3\rightarrow 2$ is greatly preferred and results in a streamlined FBPS generator,
it is crucial to extend the method to $2\rightarrow 1$ clustering jet algorithms so
these FBPS generators can be used for predictions of measured observables,
In ref.~\cite{Giele:2015sva} the first step in that direction was taken for lepton collider 
multi-jet observables using the experimental jet algorithms. 
Lepton collider observables remove most of the complexities arising in hadron collider jet
observables, most importantly the initial state hadronic radiation.
In this paper we extend the method to hadron colliders, taking into account jet masses and
initial state radiation. The FBPS generator constructed in this paper can be used to make
predictions for the LHC experiments using the experimental jet algorithms. 
Before being able to formulate such an improved method, some issues need to be addressed.

For final state radiation, defined as any partonic \brem radiation
clustered into a jet, the jet algorithm will construct jet momenta
which at higher orders are no longer massless.
As a result the massless Born jets cannot be associated with the higher order
jets. However, we need to consider the jet event as semi-inclusive as far
as the partonic contributions are concerned. In other words
we have to integrate over all partonic configurations and consequently, jet masses, resulting
in the final state \brem contribution to the probabilistic weight of the observed $V+$ jet event.
In order to accomplish this, the higher order events with massive
jets need to be projected onto the Born event with massless jets.
This results in opaque jets without any internal structure as far as perturbative QCD is concerned.
For hadron colliders the projection choice
is obvious, one leaves the transverse momentum of the jet, $\vec p_T^{\mbox{\tiny JET}}$, and
the jet rapidity, $\eta_{\mbox{\tiny JET}}$, invariant under the projection. This means
that no predictions can be made which depend on the jet masses, as it is integrated over.
Observables are constructed from the Born momenta with higher order corrections changing the
event weight but {\it not} the event kinematics.

For initial state radiation, defined as partonic \brem radiation not associated with the
jets, the issue is in principle the same but in practice a bit more subtle.
Any event has to observe momentum conservation.
That is, summing the momenta of the final state objects
together with the momenta of the unclustered radiation gives
a perfect momentum balance (modulo detector generated restrictions).
To obtain a physical result we
need to integrate over the unclustered partonic \brem radiation.
Many avenues to proceed are possible and in this paper we will consider two options for
projecting the event onto to Born event.
The first option considers all partonic \brem radiation as final state radiation,
terminating the jet clustering when a preset number of clusters is reached. 
In the case of $V+1$ jet all partonic radiation would be combined in a single cluster.
The second option follows the traditional jet cone size approach. The transverse momentum of
the unclustered parton(s) is taken into account by adding it back to one of the final state
objects. As a result the final state objects are balanced in transverse momentum
and hence can be associated with a LO momentum configuration. Naively, one would think this would
make the event weight sensitive to the initial state radiation. However, the opposite is true.
Not accounting for the initial state radiation into the final state objects lead to instabilities
in the perturbative predictions. For example the inclusive Vector Boson transverse momentum
distribution is sensitive to the initial state radiation because this radiation is not accounted
for~\cite{Corcella:1999gs}. As a result a fixed order prediction of this observable will fail.

With the above methods of projecting the partonic \brem events onto the Born phase space, we will
factorize in Section~\ref{FBPS} the \brem higher order phase space
into the Born phase space times the \brem phase space. This adds
a single 4-momentum integration per added \brem parton.
The initial and final state FBPS generators will, in Section~\ref{EventGenerator},
be combined into a single event generator.
While most of the techniques in this paper are applicable to $PP\rightarrow V+$ jets and to
a lesser extend $PP\rightarrow$ jets, for the more explicit implementation and examples
we restrict ourselves to $PP\rightarrow V+1$ jet. 
The extension to multiple jets is a straightforward generalization. 
Finally, the results are summarized in Section~\ref{Conclusions}.

\section{The Forward Branching Phase Space Generator}\label{FBPS}

In this section we factorize the \brem phase space into a Born phase space
times an integral over the \brem parton momenta. The Born phase space is defined
by a preset projection prescription such as a jet algorithm. By construction,
restricting the phase space integration to a single Born phase space point 
will integrate out the \brem contribution to that
particular Born event. This enables one to calculate the radiative corrections
order-by-order in perturbation theory to any observable constructed from these
fixed Born momenta.

To obtain the factorization of the phase space in a Born phase space times the
contribution from the \brem\!\!, 
we adopt the Forward Branching Phase Space (FBPS) approach~\cite{Giele:2011tm,Giele:2015sva}.
This method is firmly based on the methods used in parton showers where one generates new
radiation starting from a Born event~\cite{Buckley:2011ms}. We imprint on the branching the projective
constraints so the \brem integrates over the parton configurations reconstructing to
the fixed Born event.

A FBPS generator starts from a Born phase space generator producing the momenta
of all observable objects from which observables are constructed.
For the process of Vector Boson production plus $n$ jets,
$P(\hat p_a)P(\hat p_b)\rightarrow p_V(\hat Q)+J_1(\hat p_1)+\cdots+ J_n(\hat p_n)$,
the Born phase space is given by
\beq\label{BornPS}
d\Phi(\hat p_a\hat p_b;\hat Q,\{\hat p\}_n)=\left(\prod_{i=1}^n
  \frac{d \hat p_i}{(2\pi)^3}\delta(\hat p_i^2)\right) \frac{d
  \hat Q}{(2\pi)^3}\delta(\hat Q^2-M_V^2)
\delta(\hat p_a+\hat p_b-\hat p_1-\cdots-\hat p_n-\hat Q)\ ,
\eeq
where $\hat p_a$ and $\hat p_b$ are the incoming parton momenta, $\hat Q$ the vector
boson momentum, $M_V$ the vector boson mass and $\{\hat p\}_n=\{\hat p_1,\hat p_2\ldots,\hat p_n\}$
the set of $n$ jet momenta.
By generating an additional parton we obtain the \brem phase space which will have a generic
FBPS form. For a NLO generator this is given by
\beqa
d\Phi(p_ap_b;Q,\{p\}_{n+1})&=&d\Phi(\hat p_a\hat p_b;\hat Q,\{\hat p\}_n)\\
&\times&\left(\frac{d p_{n+1}}{(2\pi)^3}\delta(p_{n+1}^2)\right)
J(\{\hat p\}_n, p_{n+1})\times\delta(M(\{\hat p\}_n\rightarrow\{p\}_{n+1}))\ ,  \nonumber
\eeqa
where we can choose a map generating the \brem momenta from the Born momenta
and the Jacobian $J(\{\hat p\}_n, p_{n+1}\})$ needs to be calculated.
The projective jet algorithm is given by the inverse map
$M^{-1}(\{\hat p\}_n\leftarrow\{p\}_{n+1})$.
A minimum requirement for the map is that $\hat Q^2=Q^2$. Furthermore, it
is advantageous to separate initial state radiation from final state
radiation. One can iterate this factorization trivially to generate multiple \brem
particles relevant for beyond NLO calculations.

The first step is to define the cluster map, which is as close to the
experimentally used object reconstruction as possible. 
Any infrared safe jet algorithm will suffice, including the family of
$k_T$-jet algorithm.
For final state radiation a parton branches into two
partons generating a jet mass. By keeping the transverse momentum, 
$\vec{p}_T=(p_x,p_y)$, and the rapidity, 
$\eta=\frac{1}{2}\log\left(\frac{E+p_z}{E-p_z}\right)$, 
of the jet invariant under the branching one can
integrate over the jet mass.
Obtaining the massless Born jet momentum is a numerically straightforward
procedure. For a jet momentum with mass, we rewrite it in coordinates of mass,
transverse momentum and rapidity, set the mass to zero and translate it back in a now
massless jet with the same transverse momentum and rapidity:
$p_\mu=(p_x,p_y,p_z,E)\rightarrow (m_j,\vec p_T,\eta)\rightarrow (0,\vec p_T,\eta)\rightarrow
\hat p_\mu=(\hat p_x,\hat p_y,\hat p_z,\hat E)$.
For the initial state radiation one generates unclustered particles
(or particles ``clustered with the beam jet'') creating a
transverse momentum imbalance in the final state object momenta.
Here one can choose to absorb the generated unclustered momenta in
the vector boson momentum or alternatively
keep clustering \brem momenta into jets until the required number
of remaining clusters is obtained by not defining a jet cone size.
Using this type of projection allows one to calculate the differential cross sections
$d^{(4)}\sigma/d Q\{d\hat{\vec p}_T^{(i)} d\hat\eta_i\}_{i=1}^n$ point-by-point
in the distribution without the need to introduce histograms.
Specifying a single point in the distribution gives an unique
Born event with fixed momenta, while higher order corrections add a $K$-factor
calculable order-by-order in perturbation theory.

\subsection{The Born Phase Space Generator}

We begin by formulating the Born phase space generator in a convenient manner and 
introduce the notation and conventions that we will use for the remainder of the paper.
In order to be able to constrain the rapidity of the vector boson, $\hat\eta_Q$,
the transverse momenta, $\hat p_T^{(i)}$, rapidity,
$\hat\eta_i$, and the azimuthal angle, $\hat\phi_i$, of the jets,
we will make these kinematic variables explicit integration variables.

We use the convention that hatted variables
are for the Born event and unhatted variables are for the branched \brem events.
The momenta are parameterized as follows
\beqa\label{memrecon}
\hat p_a&=&\frac{1}{2}\left(0,0,\hat x_a,\hat x_a\right)\nonumber\\
\hat p_b&=&\frac{1}{2}\left(0,0,-\hat x_b,\hat x_b\right)\nonumber\\
\hat p_{\mu}^{(i)}&=&\hat p_T^{(i)}\left(\sin\hat\phi_i,\cos\hat\phi_i,\sinh\hat\eta_i, \cosh\hat\eta_i\right)\nonumber\\
\hat Q_{\mu}&=& \left(\hat{\vec q}_T,\hat\alpha_T\sinh\hat\eta_q, \hat\alpha_T\cosh\hat\eta_q\right)\ \mbox{where}\
\hat\alpha_T=\sqrt{\hat q_T^2+M_V^2},
\eeqa
where $\hat p_{a,b}$ are the the incoming particle momenta with parton fractions $x_{a,b}$, the jet
momenta are given by $\hat p_{\mu}^{(i)}$, and the vector boson momentum is given by $\hat Q_\mu$.
Starting from the Born phase space, Eq.~\ref{BornPS}, we perform the following transformations:
\beqa
\frac{d\hat p_i}{(2\pi)^3}\delta(\hat p_i^2)&=&\frac{d\hat{\vec p_i}}{2(2\pi)^3 E_i}
=d\hat p_T^{(i)}d\hat\eta_id\hat\phi_i\left(\frac{\hat p_T^{(i)}}{2(2\pi)^3}\right)\nonumber\\
\frac{d\hat Q}{(2\pi)^3}\delta(\hat Q^2-M_V^2)&=&\frac{d\hat{\vec Q}}{2(2\pi)^3 E_q}
=\frac{d\hat{\vec q}_T d\hat\eta_q}{2(2\pi)^3}\ \mbox{(where we used $d\hat q_z=\hat E_q d\hat\eta_q$)}\nonumber\\
\delta^{(4)}\left(\hat p_a+\hat p_b-\hat p_1-\cdots-\hat p_n-\hat Q\right)&=&
\delta^{(2)}\left(\sum_{i=1}^n\hat{\vec p}_T^{(i)}+\hat{\vec q}_T\right)\nonumber\\
&\times&\delta\left(\frac{1}{2}\sqrt{S}(\hat x_a+\hat x_b)
-\sum_{i=1}^n \hat p_T^{(i)}\cosh\hat\eta_i-\hat\alpha_T\cosh\hat\eta_q\right)\nonumber\\
&\times&\delta\left(\frac{1}{2}\sqrt{S}(\hat x_a-\hat x_b)
-\sum_{i=1}^n \hat p_T^{(i)}\sinh\hat\eta_i-\hat\alpha_T\sinh\hat\eta_q\right)\ .\nonumber\\
\eeqa
After integrating $\hat{\vec q}_T$, $\hat x_a$, $\hat x_b$ over the Kronecker $\delta$-functions, we
obtain
\beq
d\hat x_ad\hat x_bd\Phi(\hat p_a\hat p_b;\hat Q,\{\hat p\}_n)=
\frac{1}{(16\pi^3)^{n+1}}\frac{2}{S}
\left(\prod_{i=1}^nd\hat p_T^{(i)}d\hat\eta_id\hat\phi_i\times\hat p_T^{(i)}\right)
\times d\hat\eta_q\times\Theta(1-\hat x_1)\Theta(1-\hat x_2)\ ,
\eeq
with
\beqa
\hat{\vec q}_T&=&-\sum_{i=1}^n\hat{\vec p}_T^{(i)}\nonumber\\
\hat x_a&=&\frac{1}{\sqrt{S}}\left(\hat\alpha_T e^{\eta_q}+\sum_{i=1}^n\hat p_T^{(i)} e^{\hat\eta_i}\right)
\nonumber\\
\hat x_b&=&\frac{1}{\sqrt{S}}\left(\hat\alpha_T e^{-\eta_q}+\sum_{i=1}^n\hat p_T^{(i)} e^{-\hat\eta_i}\right)\ ,
\eeqa
and the constraints for a physical event, $\hat x_{a,b}\leq 1$, given by the $\Theta$-functions.

\subsection{The Final State FBPS Generator}
\begin{figure}
\centering
\includegraphics[width=7.0cm]{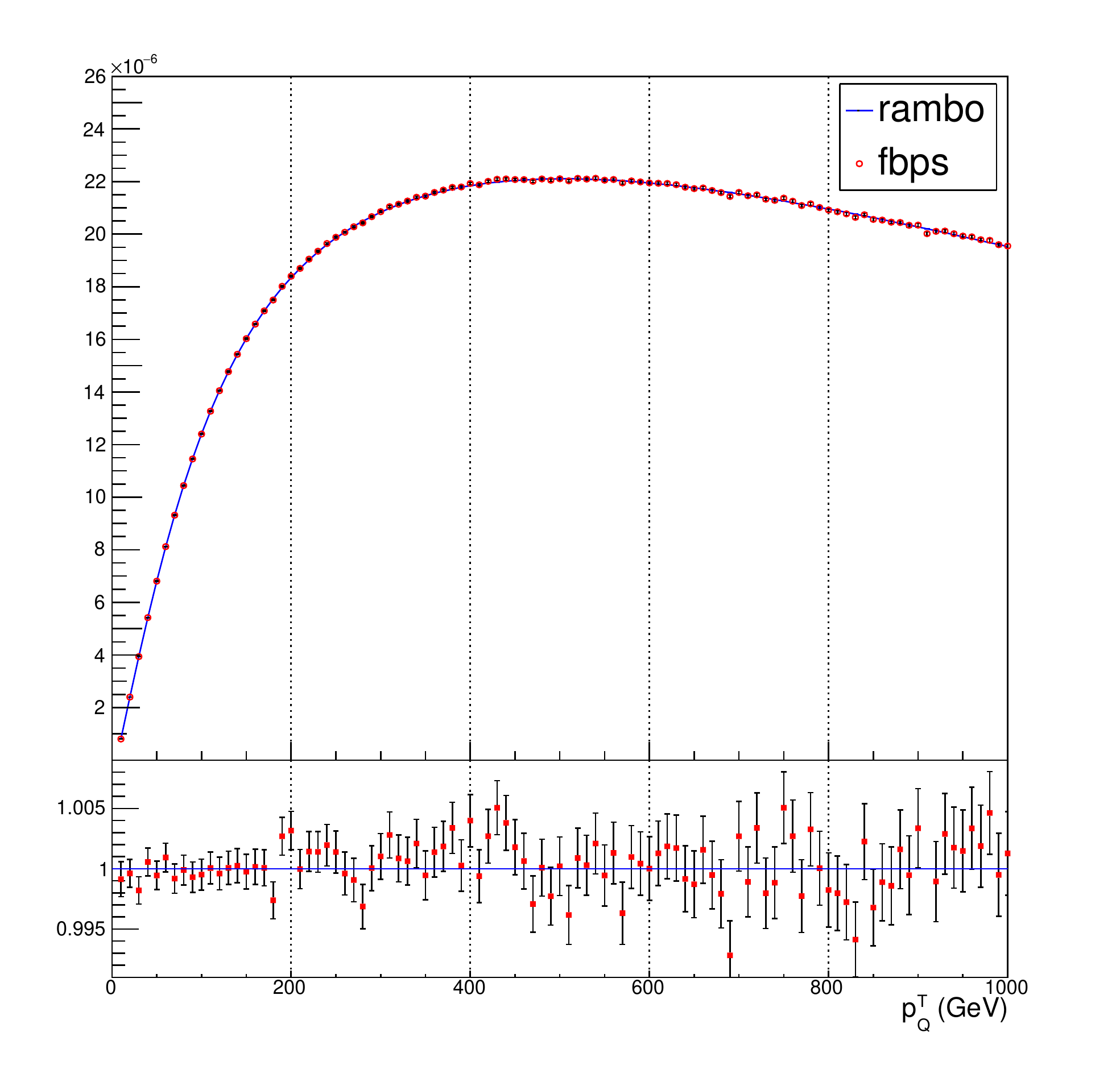}
\includegraphics[width=7.0cm]{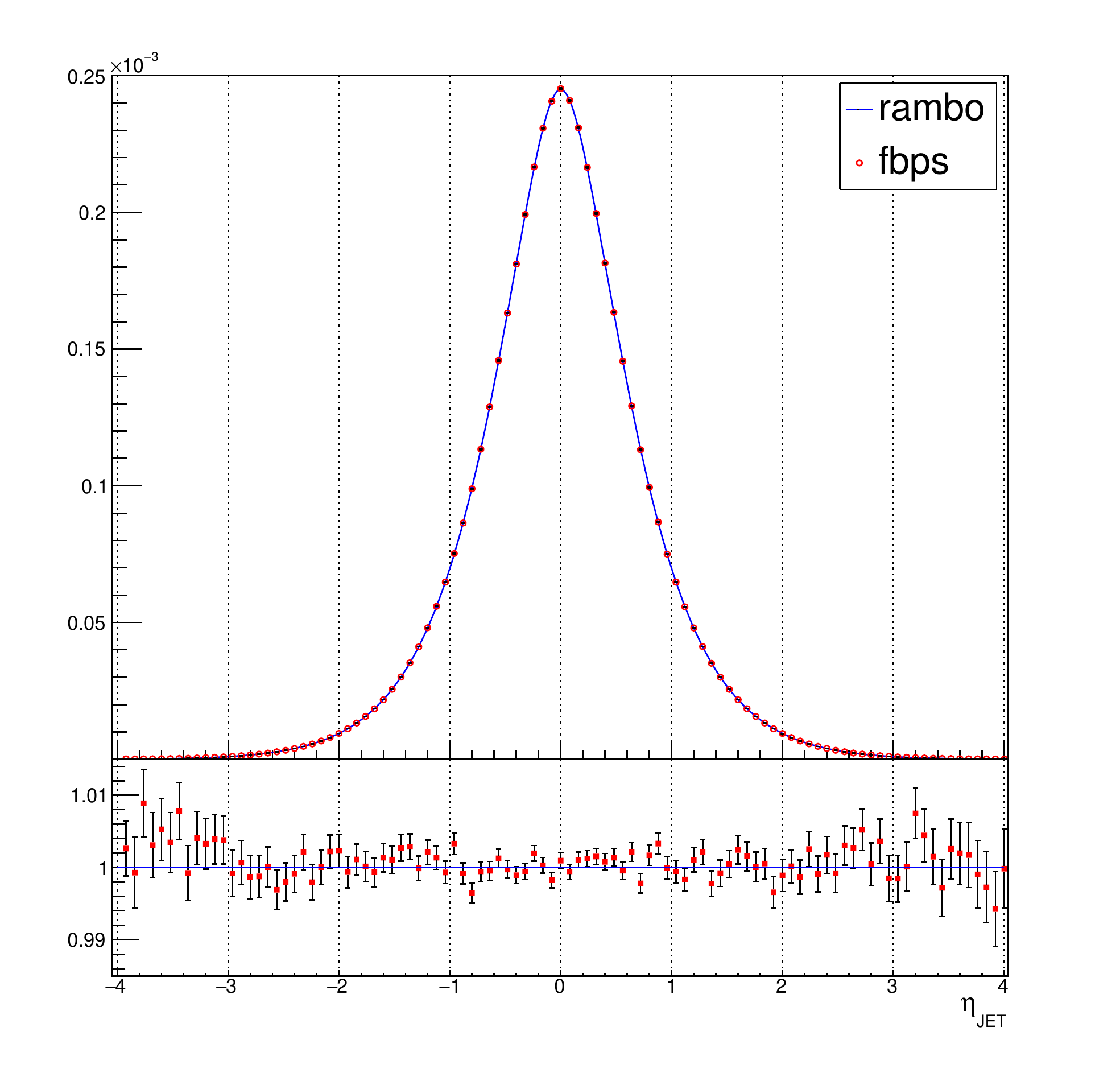}
\caption{\label{Compfinal}
A comparison between RAMBO and the FBPS generator of Eq.~\ref{FBPSfinal}. The left
graph compares the vector boson transverse momentum, 
while the right graph compares the jet rapidity.
For Rambo the $V+$ 2 particle phase space was generated and a jet algorithm
applied to obtain a $V+1$ jet inclusive final state. For the FBPS generator
the observables are the Born momenta, while the phase space weight is re-weighted
by the \brem event(s).
}\end{figure}
The final state generator begins with a $V +$ jet(s) Born event.
One of the massless Born jets is split into two partons
generating a massive jet. To project back onto the massless Born jet we
scale the longitudinal momentum, $\vec p_{12}^L$ of the massive jet,
leaving invariant both
the transverse momentum and the rapidity of the jet
\beq
p_a+p_b-Q-p_1-p_2-P=p_{ab}-Q-p_{12}-P=(p_{ab}+\alpha\,p_{12}^L)-Q-(p_{12}+\alpha\, p_{12}^L)-P
=\hat p_{ab}-\hat Q-\hat p_J-\hat P\ ,
\eeq
where $\hat Q=Q$, $\hat p_{ab}=\hat p_a+\hat p_b=p_a+p_b+\alpha\,p_{12}^L$,
$\hat p_J=p_{12}+\alpha\,p_{12}^L$, $p_{12}^L=(p_1+p_2)_L=(0,0,(p_1)_z+(p_2)_z,E_1+E_2)
, P=\hat P=p_3+\cdots+p_n$
and $\alpha$ is given by the
constraint $\hat p_J^2=0$. Note that $\hat p_a$ and $\hat p_b$ are reconstructed from
$\hat p_{ab}$ using $\frac{1}{2}\sqrt{S}\times\hat x_{a,b}=\hat E_{ab}\pm\left(\hat p_z\right)_{ab}$
in the momenta reconstruction of Eq.~\ref{memrecon}.

To derive the FBPS generator we first encode the branching in a decomposition of unity
\beq
1=2\sqrt{(p_{12}^L)^2(p_{12}^T)^2}\int d \alpha\,
d\hat p_J\,\delta(\hat p_J^2)\delta(\hat p_J-p_{12}-\alpha p_{12}^L)\ ,
\eeq
insert it into the 3-particle phase space generator
\beq
d\Phi_3(p_a,p_b;Q,p_1,p_2)=\frac{d\,Q}{(2\pi)^3}\frac{d\,p_1}{(2\pi)^3}
\frac{d\,p_2}{(2\pi)^3}\ \delta(Q^2-M_V^2) \delta(p_1^2)
\delta(p_2^2)\delta(p_a+p_b-Q-p_1-p_2)\ ,
\eeq
and integrate $p_2$ over $\delta(\hat p_J-p_{12}-\alpha\, p_{12}^L)$ and
subsequently integrate $\alpha$ over $\delta(p_2^2)$ to obtain the desired FBPS generator
\beq\label{FBPSfinal}
d\Phi_3^{\mbox{\tiny FINAL}}(p_a,p_b;Q,p_1,p_2)
=d\Phi_2(\hat p_a,\hat p_b;\hat Q,\hat p_J)\times\left[\frac{d\,p_1}{(2\pi)^3}\
  \delta(p_1^2)\right]\times J(\hat p_J,p_1)\ .
\eeq
The momenta are given by $p_2^T=\hat p_J^T-p_1^T$, $p_2^L=\beta_+\hat p_J^L-p_1^L$,
$p_{ab}=\hat p_{ab}-(1-\beta_+)\,\hat p_J^L$, $Q=\hat Q$ and the Jacobian is given by
\beq
J(\hat p_J,p_1)=\left|\frac{2}{1-\beta_-/\beta_+}\right|\times
\sqrt{\frac{(\hat p_J^T)^2}{(\hat p_J^L)^2}}
=\left|\frac{2}{1-\beta_-/\beta_+}\right|\ ,
\eeq
where
\beq
\beta_\pm=\frac{(\hat p_J^L\cdot p_1^L)\pm\sqrt{(\hat p_J^T)^4+2 (\hat p_J^L)^2
    (\hat p_J^T\cdot p_1^T)+ (\hat p_J^L\cdot p_1^L)^2}}{(\hat p_J^L)^2}\ .
\eeq

The upper limit on the integration over $p_1$ is determined by
the condition on the parton fractions $x_{a,b}<1$. Because the
\brem is generated mostly soft and/or collinear with the jet
momentum this condition can be implemented through a veto without any
noticeable impact on the performance of the generator.

The generator is a complete phase space generator, however it is
designed to be used in a different manner.
This final state brancher is generated from a fixed Born event. By
repeatedly branching this event, one integrates over the jet mass
and obtains the final state part of the $K$-factor for this Born
event.
The FBPS generator of Eq.~\ref{FBPSfinal} is compared to the RAMBO
phase space generator~\cite{Kleiss:1985gy} in Fig.~\ref{Compfinal}.

\subsection{The Initial State FBPS Generator}

\begin{figure}
\centering
\includegraphics[width=7.0cm]{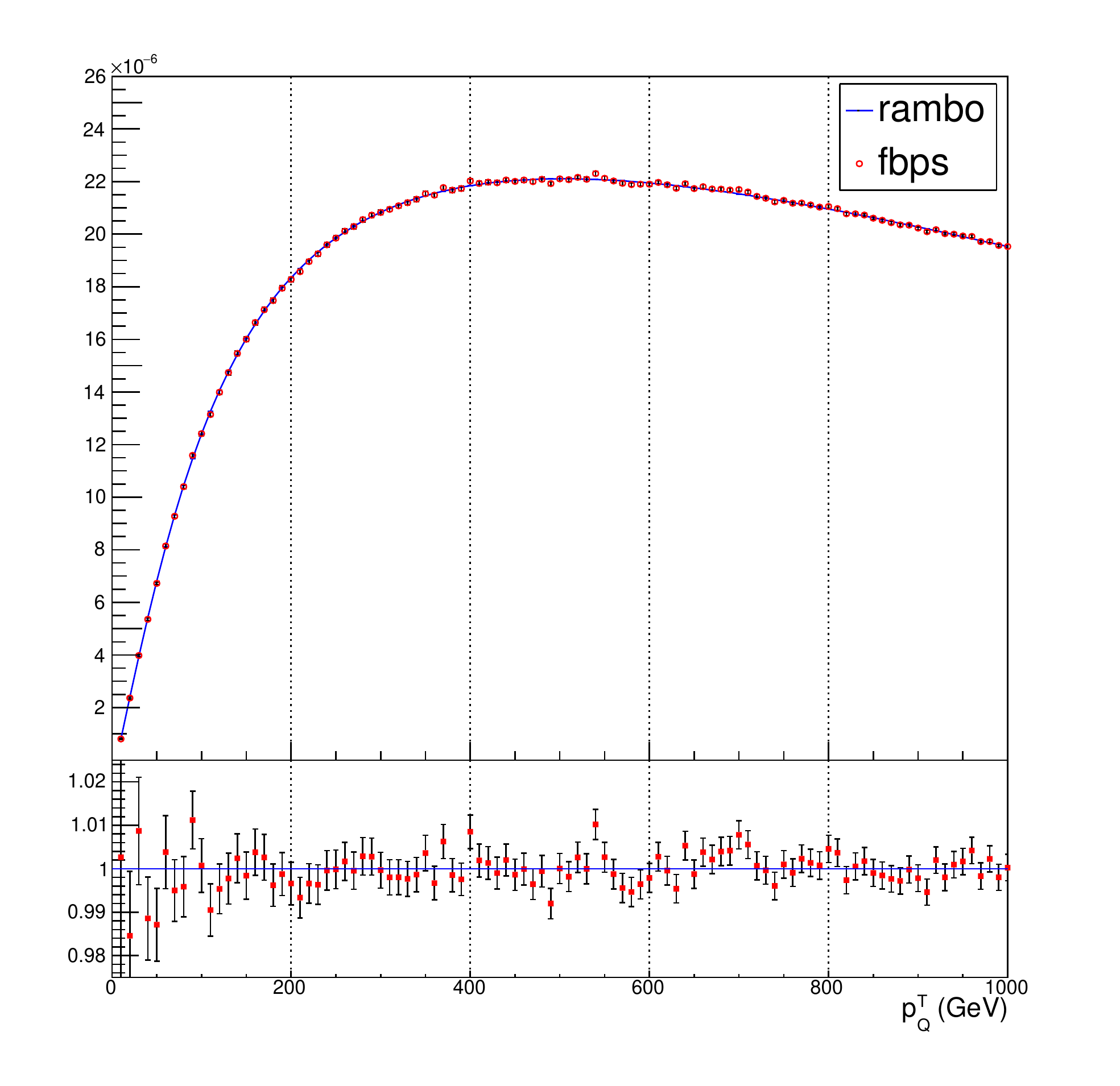}
\includegraphics[width=7.0cm]{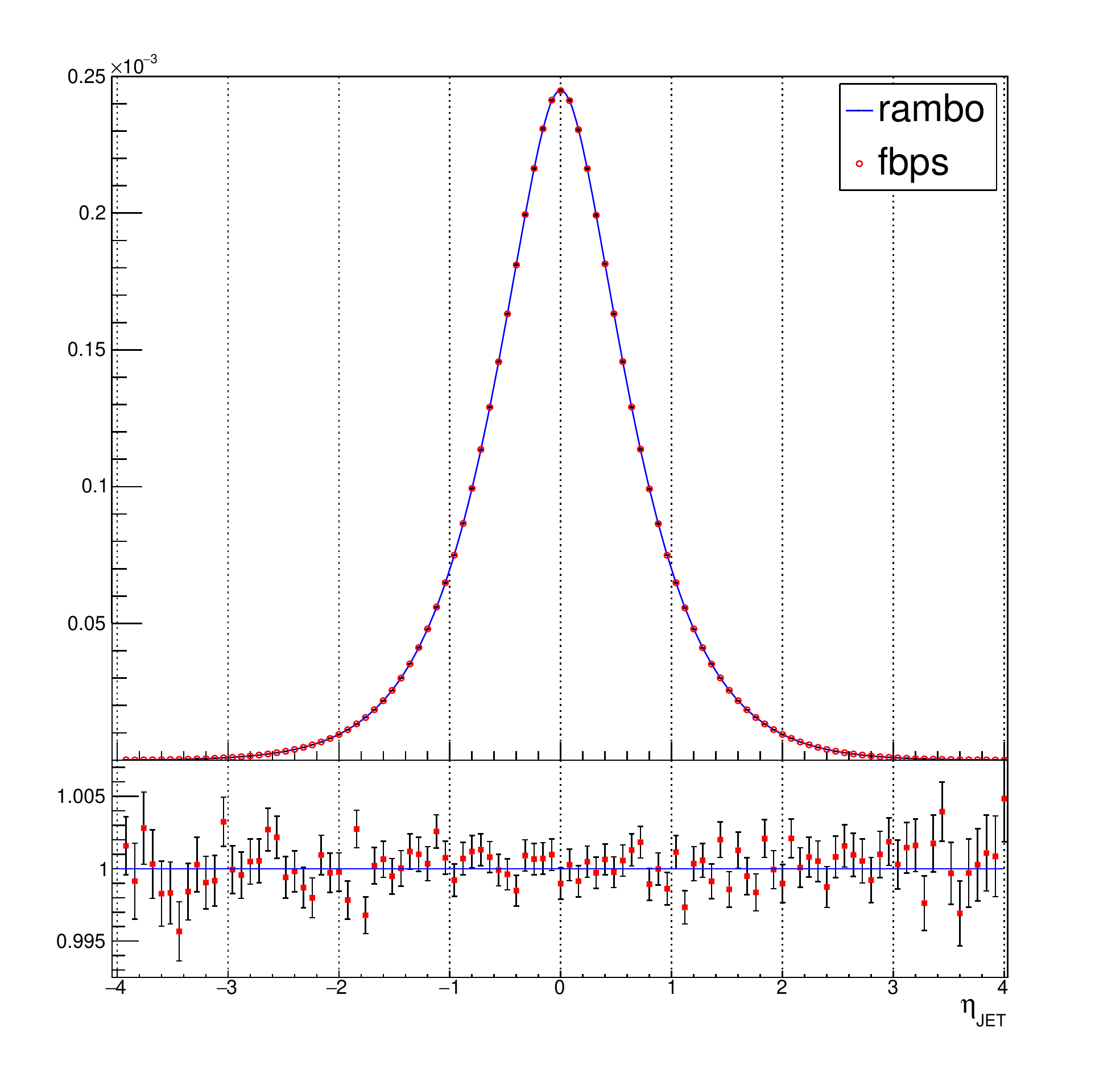}
\caption{
A comparison between RAMBO and the FBPS generator of Eq.~\ref{FBPSinit1}. The left
graph compares the vector boson $p_T$, while the right graph compares the jet rapidity.
For Rambo the $V+$ 2 particle phase space was generated and a jet algorithm
applied to obtain a $V+1$ jet exclusive final state. For the FBPS generator
the observables are the Born momenta, while the phase space weight is re-weighted
by the \brem event(s).
}
\label{Compinit1}
\end{figure}
The initial state brancher is more complicated due to the fact that
the extra parton generated is part of the beam jet and not added to
the final state jets. As a consequence the transverse
momentum generated by the branching momentum will have to be balanced by vector
boson or absorbed into the jet.
At higher orders this generates events with transverse momentum of the vector
boson below the transverse momentum cut on the jet transverse momentum.

We first derive the generator which leaves the jet momentum $p_J$
invariant. The cluster map is given by
\beq
\hat p_{ab}-\hat Q-\hat p_J=(\hat p_{ab}-\alpha p_1^L)-(\hat Q-\alpha p_1^L-p_1)-\hat p_J-p_1
=p_{ab}-Q-p_J-p_1\ ,
\eeq
where $p_J=\hat p_J$, $p_{ab}=\hat p_{ab}-\alpha p_1^L$,
$Q=\hat Q-\hat p_1^T-(1+\alpha) p_1^L$ and $\alpha$ is given by the constraint
that $Q^2=\hat Q^2=M_V^2$.

The first step is again to encode the above branching in a partition of unity
\beq
1=2\sqrt{(p_1^L)^4-2(p_1^L)^2(p_1^T\cdot Q)+(p_1^L\cdot Q)^2}\int d \alpha d\hat Q\
\delta(\hat Q^2-M_V^2) \delta(\hat Q-Q-p_1^T-(1+\alpha) p_1^L)\ ,
\eeq
and insert it in the 3 particle phase space
\beq
d\Phi_3(p_a,p_b;Q,p_J,p_1)=\frac{d\,Q}{(2\pi)^3}\frac{d\,p_J}{(2\pi)^3}\frac{d\,p_1}{(2\pi)^3}\
\delta(Q^2-M_V^2) \delta(p_J^2)\delta(p_1^2)\delta(p_a+p_b-Q-p_J-p_1)\ .
\eeq
By integrating $Q$ over $\delta(\hat Q-Q-p_1^T-(1+\alpha) p_1^L)$ and
$\alpha$ over $\delta(Q^2-M_V^2)$ one obtains the FBPS generator
\beq\label{FBPSinit1}
d\Phi_3^{\mbox{\tiny INIT,I}}(p_a,p_b;Q,p_J,p_1)=
d\Phi_2(\hat p_a,\hat p_b;\hat Q,\hat p_J)\times\left[\frac{d\,p_1}{(2\pi)^3}\
  \delta(p_1^2)\right]\times J(\hat Q,p_1)\ .
\eeq
The momenta are given by $Q_T=\hat Q_T-p_1^T$,
$Q_L=\hat Q_L-(1+\alpha) p_1^L$, $p_{ab}=\hat p_{ab}-\alpha p_1^L$,
$p_J=\hat p_J$ and the Jacobian is given by
\beq
J(\hat Q,p_1)=\sqrt{\frac{(p_1^L)^4-2(p_1^L)^2(p_1^T\cdot Q_T)
+(p_1^L\cdot Q_L)^2}{(p_1^L)^4+2(p_1^L)^2(p_1^T\cdot\hat Q_T)+(\hat p_1^L\cdot\hat Q_L)^2}}\ ,
\eeq
where
\beq
\alpha=\frac{(p_1^L\cdot\hat Q_L)-(p_1^L)^2-\sqrt{(p_1^L)^4+2(p_1^L)^2(p_1^T\cdot\hat Q_T)
+(p_1^L\cdot\hat Q_L)^2}}{(p_1^L)^2}\ .
\eeq
Because the jet transverse momentum and rapidity is invariant under the branching the
Born phase space can be generated with the given jet cuts.
The comparisons with RAMBO are shown in Fig.~\ref{Compinit1}.

\begin{figure}
\centering
\includegraphics[width=7.0cm]{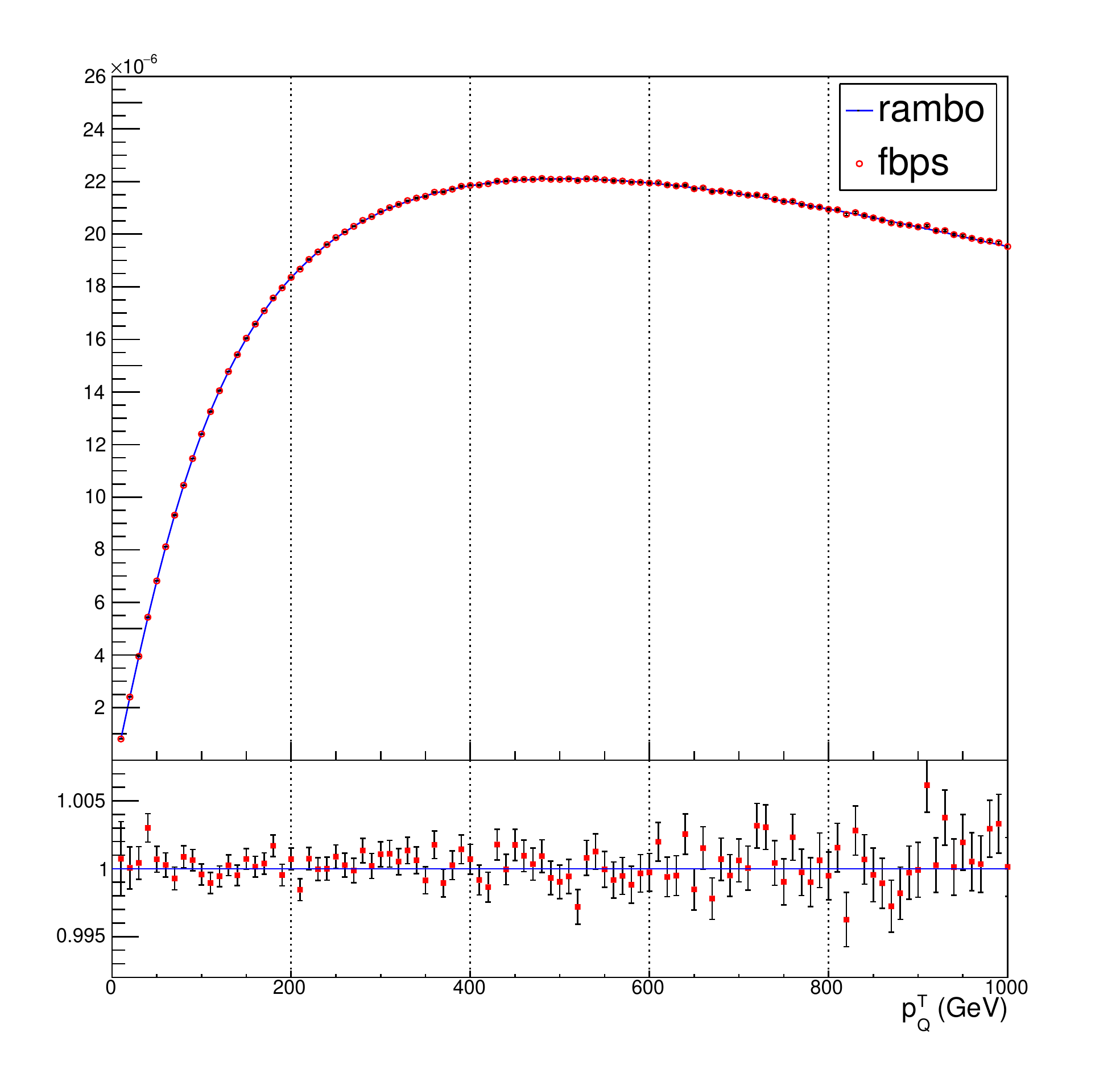}
\includegraphics[width=7.0cm]{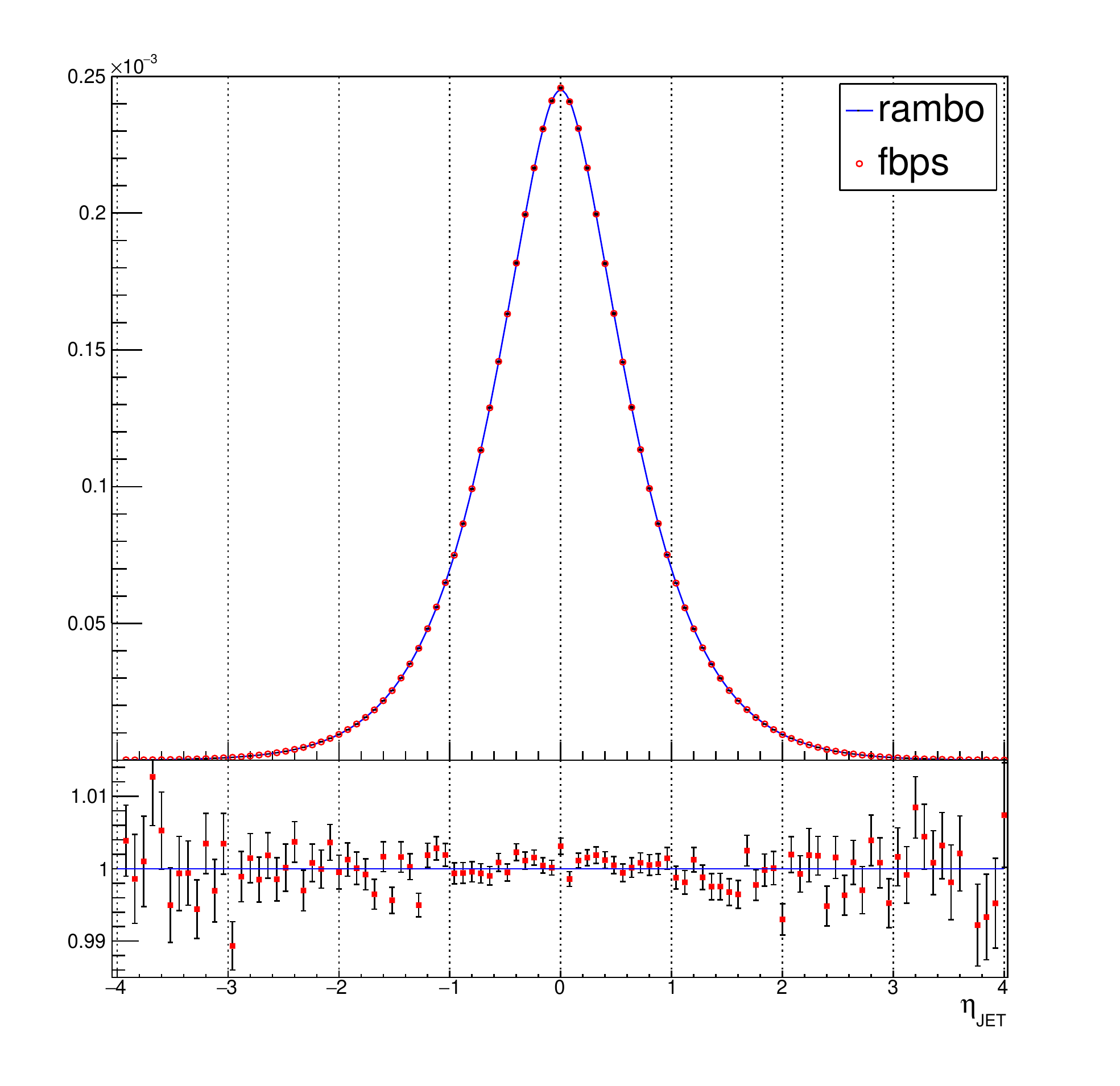}
\caption{
A comparison between RAMBO and the FBPS generator of Eq.~\ref{FBPSinit2}. The left
graph compares the vector boson $p_T$, while the right graph compares the jet rapidity.
For Rambo the $V+$ 2 particle phase space was generated and a jet algorithm
applied to obtain a $V+1$ jet exclusive final state. For the FBPS generator
the observables are the Born momenta, while the phase space weight is re-weighted
by the \brem event(s).
}
\label{Compinit2}
\end{figure}
Next we construct the generator, leaving the vector boson momentum $Q$ invariant.
The clustering is now given by
\beq
\hat p_{ab}-\hat Q-\hat p_J=(\hat p_{ab}-\alpha p_1^L)-\hat Q-(\hat p_J-\alpha p_1^L-p_1)-p_1
=p_{ab}-Q-p_J-p_1\ ,
\eeq
where $Q=\hat Q$, $p_{ab}=\hat p_{ab}-\alpha p_1^L$,
$p_J=\hat p_J- p_1^T-(1+\alpha) p_1^L$ and
$\alpha$ is given by the constraint that $ p_J^2=\hat p_J^2$.

Encoding the above branching in a partition of unity
\beq
1=2\sqrt{(p_1^L)^4-2(p_1^L)^2(p_1^T\cdot p_J^T)
+(p_1^L\cdot p_J^L)^2}\int d \alpha d\hat p_J\
\delta(\hat p_J^2) \delta(\hat p_J-p_J-p_1^T-(1+\alpha) p_1^L)\ ,
\eeq
inserting it in the 3-particle phase space,
\beq
d\Phi_3(p_a,p_b;Q,p_J,p_1)=\frac{d\,Q}{(2\pi)^3}\frac{d\,p_J}{(2\pi)^3}
\frac{d\,p_1}{(2\pi)^3}\ \delta(Q^2-M_V^2) \delta(p_J^2)
\delta(p_1^2)\delta(p_a+p_b-Q-p_J-p_1)\ 
\eeq
and subsequently
integrating $p_J$ over $\delta(\hat p_J-p_J-p_1^T-(1+\alpha) p_1^L)$
and $\alpha$ over $\delta( p_J^2)$ one obtains
\beq\label{FBPSinit2}
d\Phi_3^{\mbox{\tiny INIT,II}}(p_a,p_b;Q,p_J,p_1)
=d\Phi_2(\hat p_a,\hat p_b;\hat Q,\hat p_J)\times\left[\frac{d\,p_1}{(2\pi)^3}\
  \delta(p_1^2)\right]\times J(\hat p_J,p_1)\ .
\eeq
The momenta are given by $p_J^T=\hat p_J^T-p_1^T$,
$p_J^L=\hat p_J^L-(1+\alpha) p_1^L$, $p_{ab}=\hat p_{ab}-\alpha p_1^L$,
$Q=\hat Q$ and the Jacobian is given by
\beq
J(\hat p_J,p_1)=\sqrt{\frac{(p_1^L)^4-2(p_1^L)^2(p_1^T\cdot p_J^T)
+(p_1^L\cdot p_J^L)^2}{(p_1^L)^4+2(p_1^L)^2(p_1^T\cdot\hat p_J^T)+(p_1^L\cdot\hat p_J^L)^2}}\ ,
\eeq
where
\beq
\alpha=\frac{(p_1^L\cdot\hat p_J^L)-(p_1^L)^2-\sqrt{(p_1^L)^4+2(p_1^L)^2(p_1^T\cdot\hat p_J^T)
+(p_1^L\cdot\hat p_J^L)^2}}{(p_1^L)^2}\ .
\eeq
Because the jet gets a kick due to the emission of the extra parton,
we cannot impose the jet cuts on the Born generator.
The comparisons with RAMBO are shown in Fig.~\ref{Compinit2}.

\subsection{Kinematics}

The above FBPS generators require us to generate the massless \brem
particle $p_1$. One method is to utilize the jettiness variables
which are equivalent to normalized Sudakov 
parameters~\cite{Stewart:2010tn,Boughezal:2015eha,Campbell:2017hsw,Campbell:2286381}. 
For the initial state brancher this is quite straightforward
\beq
d p_1\delta( p_1^2)=\frac{1}{2}d p_1^T d\eta_1
d\phi_1\times p_1^T=\frac{1}{4}d\tau_a d\tau_b d\phi_1\ ,
\eeq
with
\beq
\tau_{a/b}=( p_1\cdot p_{a/b})/E_{a/b}=p_1^T\exp{(\pm\eta_1)}\ .
\eeq
Given the integration variables $\tau_a$, $\tau_b$ and $\phi_1$
we can reconstruct the \brem momentum
\beq
p_1^T=\sqrt{\tau_a\tau_b};\
\eta_1=\frac{1}{2}\log\left(\frac{\tau_a}{\tau_b}\right)\ .
\eeq

This needs to be generalized to the final state brancher.
To define the jettiness of the
\brem particle we choose the born jet momentum $\hat p_J$ and one of the
born beam momenta, $\hat p_{a/b}$. Next, we apply a Sudakov decomposition~\cite{Sudakov:1954sw} of
the \brem momentum $p_1$ with respect to $\hat p_J$ and $\hat p_{a/b}$
\beq
p_1^\mu=\tau_J\frac{n_1^\mu}{(n_1\cdot
n_2)}+\tau_{a/b}\frac{n_2^\mu}{(n_1\cdot n_2)}+p_T^{\mu}\ ,
\eeq
with
\beq
n_1^\mu=\frac{\hat p_J^\mu}{\hat E_J};\ n_2^\mu=\frac{\hat p_{a/b}^\mu}{\hat E_{a/b}};\
n_1^2=n_2^2=(p_T\cdot n_{1/2})=0\ .
\eeq
By demanding that $p_1^2=0$ we find
\beq
p_T^2=2\frac{\tau_J\tau_{a/b}}{(n_1\cdot n_2)};\
(p_1\cdot n_1)=\frac{(p_1\cdot\hat p_J)}{\hat E_J}=\tau_J;\
(p_1\cdot n_2)=\frac{(p_1\cdot\hat p_{a/b})}{\hat E_{a/b}}=\tau_{a/b}\ .
\eeq

To construct $p_T^\mu$ we choose two space-like momenta
$n_{3/4}^\mu$ with the properties $(n_{1/2}\cdot n_{3/4})=0$ and
$n_{3,4}^2=-1$, then
\beq
p_T^\mu=\sqrt{p_T^2}\left(\sin\phi\times n_3^\mu+\cos\phi\times n_4^\mu\right)\ .
\eeq
One possible construction of $n_{3/4}^\mu$ is given by
\beq
\hat n_3^\mu=\delta_{n_1n_2g_3}^{n_1n_2\mu};\ \hat
n_4^\mu=\delta_{n_1n_2n_3g_4}^{n_1n_2n_3\mu};\ n_{3/4}^\mu=\frac{\hat
  n_{3/4}^\mu}{\sqrt{-(\hat n_{3/4}\cdot\hat n_{3/4})}}\ ,
\eeq
where $g_{3/4}^\mu$ are arbitrary reference momenta.
The integration measure is given by
\beq
d p_1\,\delta(p_1^2)=d\tau_J d\tau_{a/b} d\phi\ .
\eeq

\section{The Phase Space Generator for V+jets Cross Sections}\label{EventGenerator}

\begin{figure}
\centering
\includegraphics[width=7.0cm]{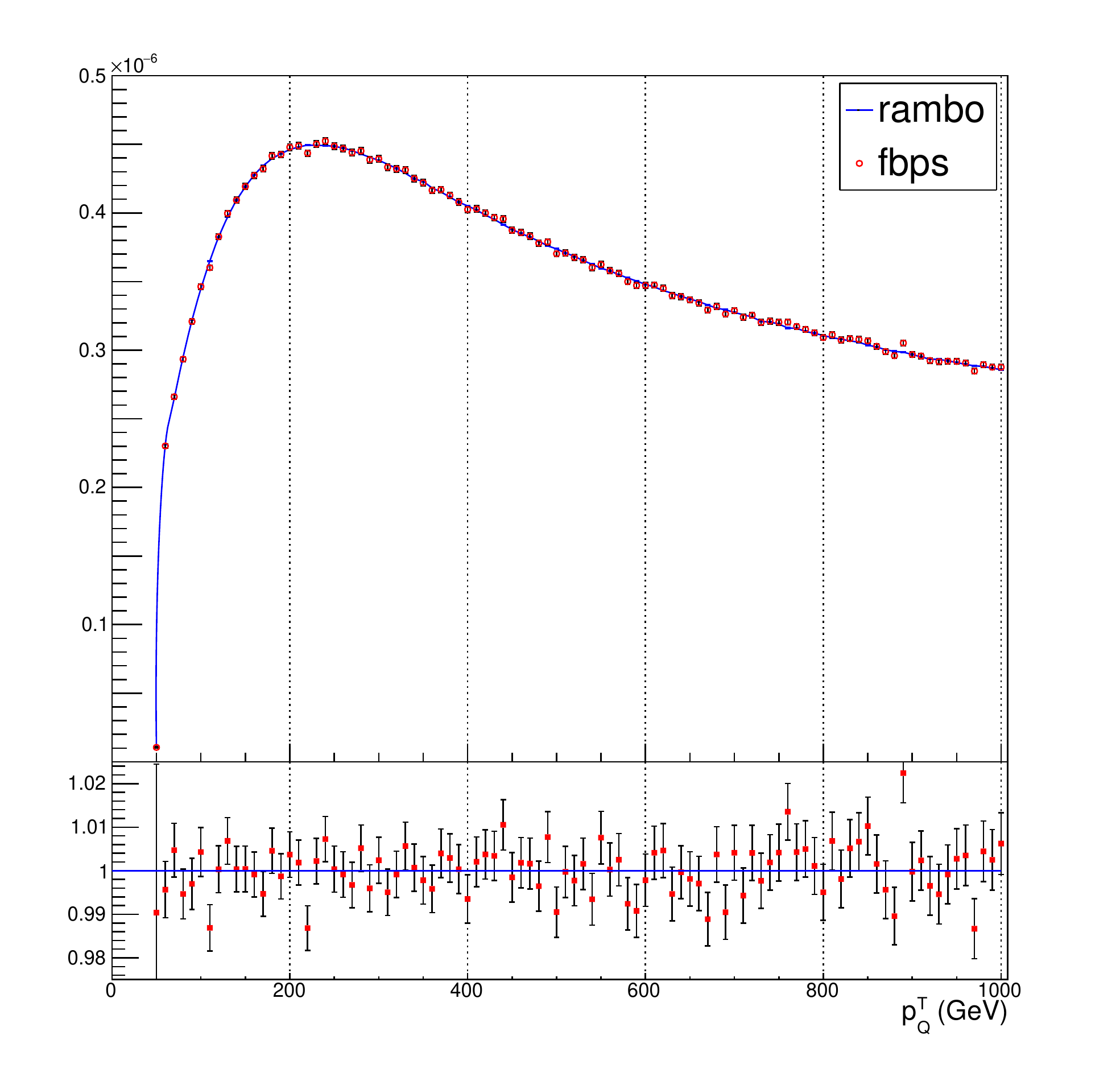}
\includegraphics[width=7.0cm]{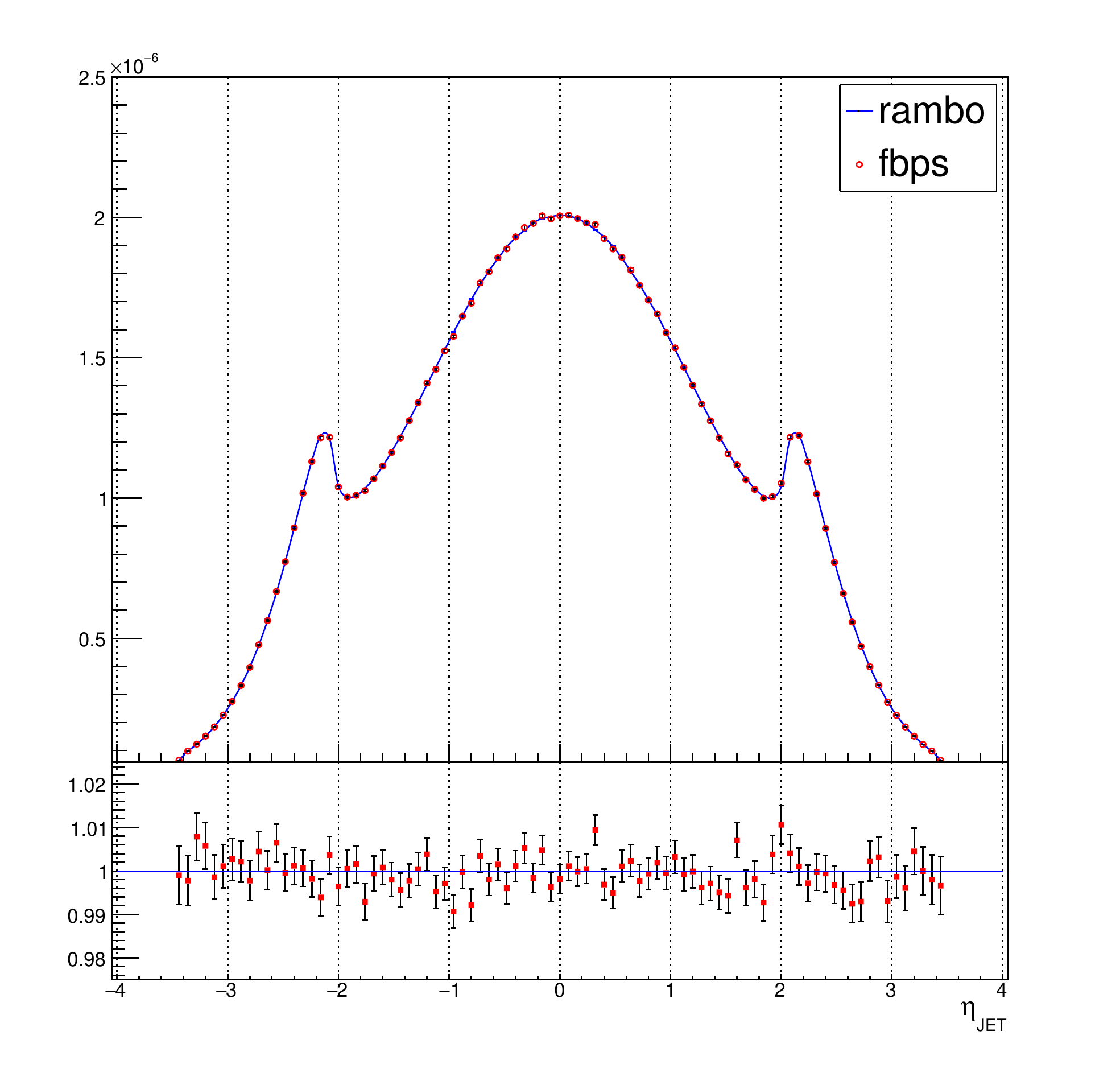}
\caption{
A comparison between RAMBO and the FBPS generator of Eq.~\ref{FBPScross}. The left
graph compares the vector boson $p_T$, while the right graph compares the jet rapidity.
For Rambo the $V+$ 2 particle phase space was generated and a jet algorithm
applied to obtain a $V+1$ jet exclusive final state. For the FBPS generator
the observables are the Born momenta, while the phase space weight is re-weighted
by the \brem event(s).}
\label{full1}
\end{figure}

The final step is to construct a cross section phase space generator.
This requires combining the initial state and final state FBPS generators.
The introduction of a jet algorithm will do this by partitioning 
the phase space into individual sectors, one for each parton. 
The phase space for the two sectors associated
with the initial state partons will use an initial state FBPS generator.
While each sector associated with a jet will use the final state
FBPS generator.

The sectors are determined by the distance measures
between the partons/clusters given by the jet algorithm.
For the $k_T$-jet family of jet algorithms~\cite{Catani:1993hr,Cacciari:2008gp} we have
\beq
d_{ij}=\min\left((p_i^T)^{2p},(p_j^T)^{2p}\right) \times\left(\frac{\Delta_{ij}^2}{R^2}\right);\
  d_{iB}=(p_i^T)^{2p}\ ,
\eeq
where
\beq
 \Delta_{ij}^2=(\eta_i-\eta_j)^2+(\phi_i-\phi_j)^2 \ .
\eeq

For the example of $PP\rightarrow V+1$ jet we use the
jet algorithm to separate the phase space into two initial state \brem sectors and
the final state \brem sector, i.e., the 0-jet, 1-jet, and 2-jet sectors. In this NLO example the jet algorithm
simplifies significantly which in general will not be the case. 
In order to separate final state \brem from other
\brem radiation, we use the decomposition of one to isolate the final state
radiation
\beq
1=\Theta(R-\Delta_{12})+\Theta(\Delta_{12}-R)\ .
\eeq
However, the second term includes 2-jet final state contributions.
We can further decompose the second term
\beq
1=\Theta(R-\Delta_{12})+\Theta(\Delta_{12}-R)\times
\left(\Theta(p_{\mbox{\tiny min}}^T-p_1^T)+\Theta(p_1^T-p_{\mbox{\tiny min}}^T)\right)\times
\left(\Theta(p_{\mbox{\tiny min}}^T-p_2^T)+\Theta(p_2^T-p_{\mbox{\tiny min}}^T)\right)\ .
\eeq
By requiring a 1-jet exclusive final state, we can filter out the 0-jet and 2-jet contributions,
resulting in
\beq
1=\Theta(R-\Delta_{12})+\Theta(\Delta_{12}-R)\left
(\Theta(p_{\mbox{\tiny min}}^T-p_1^T)\Theta(p_2^T-p_{\mbox{\tiny min}}^T)
+\Theta(p_{\mbox{\tiny min}}^T-p_2^T)\Theta(p_1^T-p_{\mbox{\tiny min}}^T)\right)
\eeq
resulting in 3 sectors, one for each of the incoming beams and one for the final state jet.
It is straightforward to generalize this procedure to multiple jet final states.

The $PP\rightarrow V+1$ exclusive jet phase space now becomes
\beqa\label{FBPScross}
\lefteqn{
d\Phi_3^{\mbox{\tiny exclusive}}(p_a,p_b;Q,p_1,p_2)=d\Phi_3(p_a,p_b;Q,p_1,p_2)}\nonumber \\
&\times&\left[
\Theta(R-\Delta_{12})\right.+\left.\Theta(\Delta_{12}-R)\left
(\Theta(p_{\mbox{\tiny min}}^T-p_1^T)\Theta(p_2^T-p_{\mbox{\tiny min}}^T)
+\Theta(p_{\mbox{\tiny min}}^T-p_2^T)\Theta(p_1^T-p_{\mbox{\tiny min}}^T)\right)
\right]\nn \\
&=&d\Phi_2(\hat p_a,\hat p_b;\hat Q,\hat p_J)\times\left[\frac{d\,p_1}{(2\pi)^3}\
  \delta(p_1^2)\right]\nn \\ &\times&
\left[\Theta(R-\Delta_{12}) J^{\mbox{\tiny FINAL}}(\hat p_J,p_1)
\delta(M^{\mbox{\tiny FINAL}}(\{\hat p\}_2\rightarrow\{p\}_2))
\right.\nn \\ && \left.
+\Theta(\Delta_{12}-R)\Big(\Theta(p_1^T<p_{\mbox{\tiny MIN}}^T) \Theta(p_2^T>p_{\mbox{\tiny
      MIN}}^T)+ (1\leftrightarrow 2)\Big) J^{\mbox{\tiny INIT}} (\hat Q,p_1)
\delta(M^{\mbox{\tiny INIT}}(\{\hat p\}_2\rightarrow\{p\}_2))\right]\nn\ . \\
\eeqa

\begin{figure}
\centering
\includegraphics[width=7.0cm]{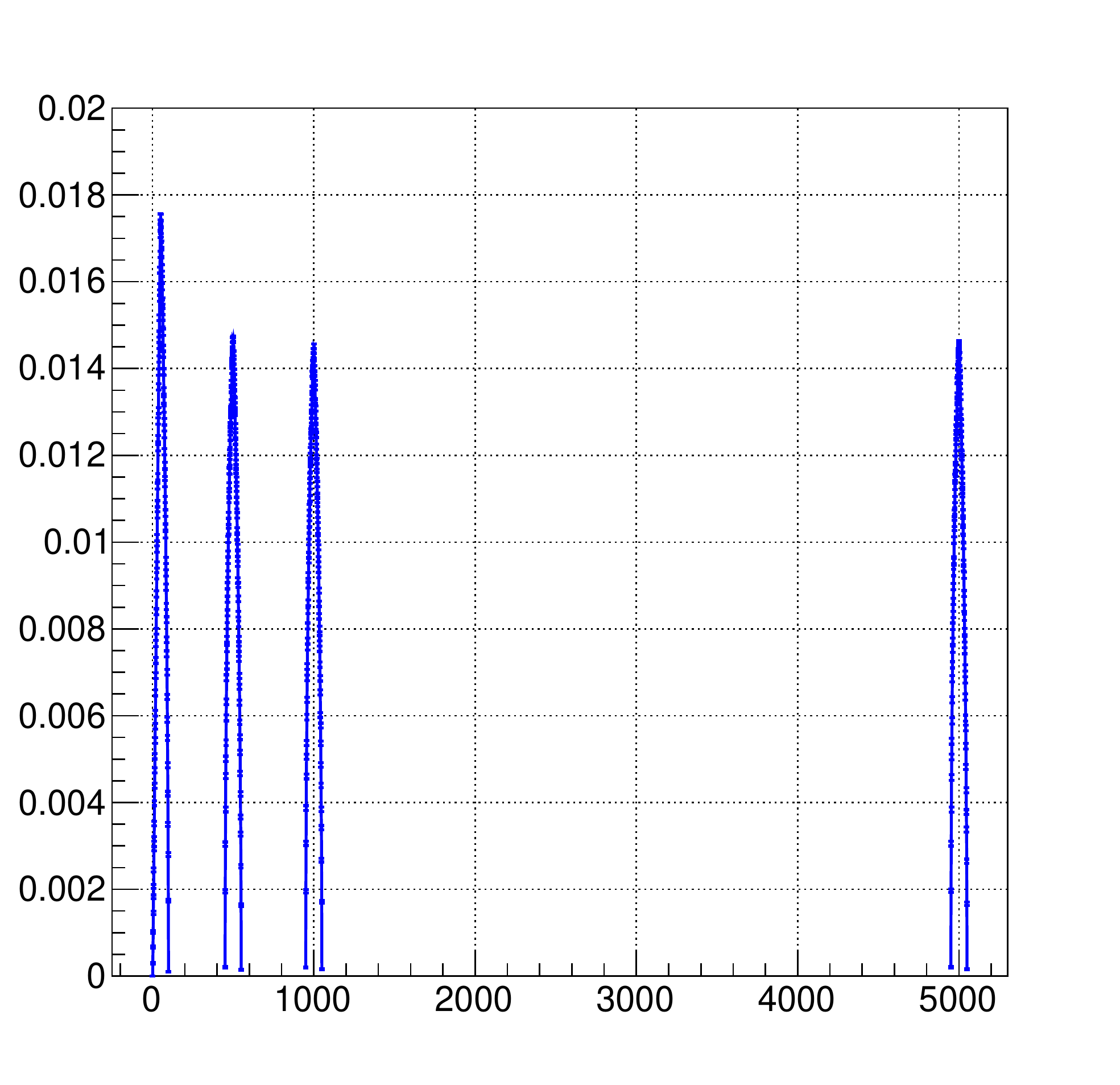}
\includegraphics[width=7.0cm]{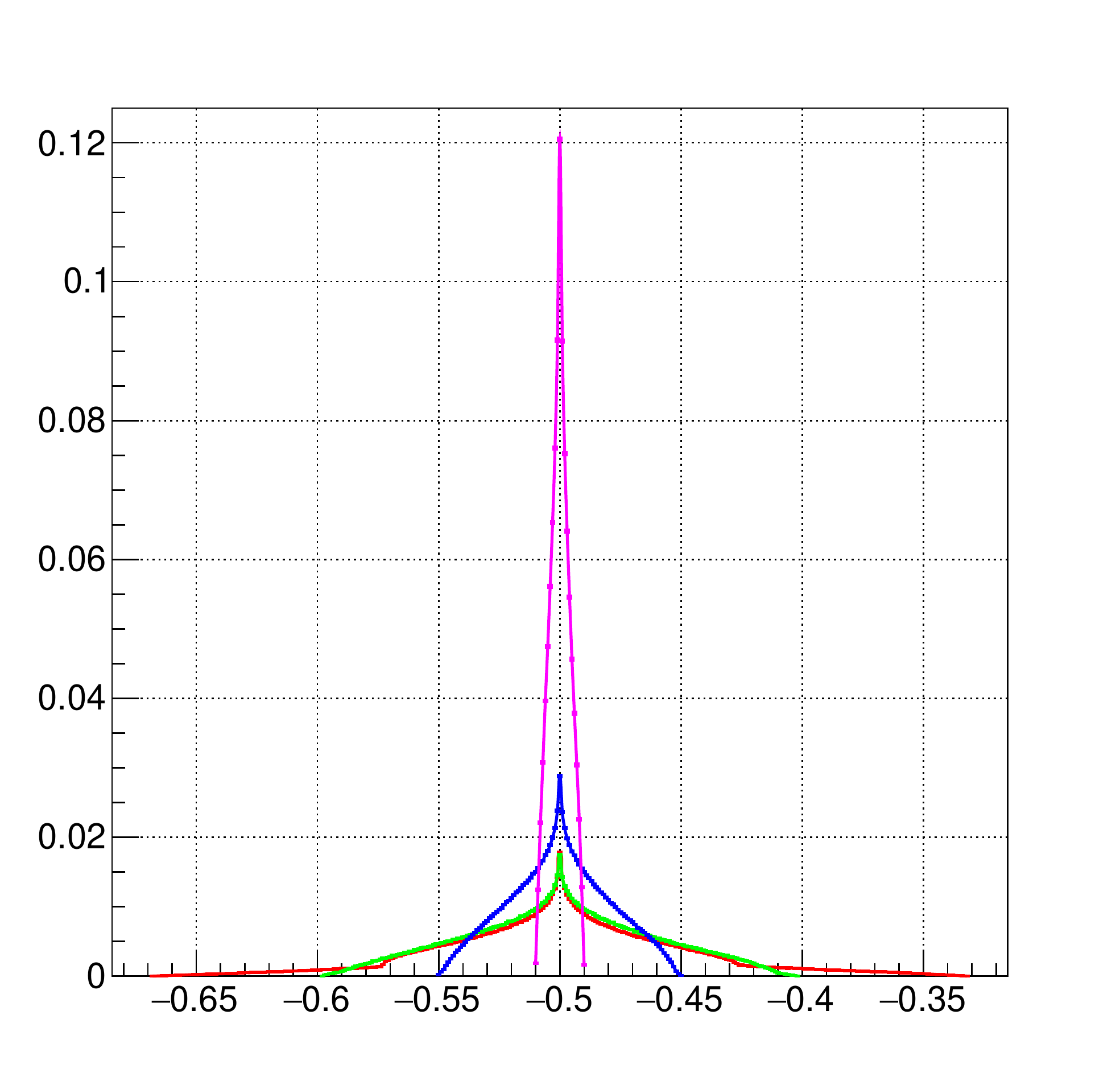}
\caption{\label{response} The transverse momentum (left) and the rapidity (right)
of the vector boson after initial state radiating for a fixed jet
momentum with transverse momentum $p^{\mbox{jet}}_T= 50, 500, 1000, 5000$
and rapidity of $0.4$ and born rapidity of the vector boson equal to $-0.5$.}
\label{initfixed}
\end{figure}
By adding the initial state transverse momentum kick 
to the vector boson we leave the jet momentum
invariant. In Fig.~\ref{full1} we show the validation of the cross section phase space
generator in Eq.~\ref{FBPScross} by comparing to the RAMBO phase space generator using
a $k_T$-jet algorithm.
To see the effect of initial state radiation we show in Fig.~\ref{initfixed} the change in
the vector boson momentum due to the initial state \brem radiation. If one chooses
to add the unclustered partonic radiation back to the vector boson momentum one would
recover fully the initiating Born event.

Alternatively, one could choose to give the 
initial state radiation transverse momentum kick to the
jet, leaving the vector boson momentum invariant. Again by balancing the transverse
momentum of the final state objects by adding the unclustered partonic transverse
momentum to the jet momentum the originating Born event is fully reconstructed.

\section{Conclusions}\label{Conclusions}

The use of a FBPS generator offers many improvements for higher order calculations
compared to conventional phase space generators. These all stem from the property
that the \brem is generated from an originating Born event with the observables
calculated from the Born event momenta. This feature provides us the ability to define single event weights beyond Leading Order.

From a numerical viewpoint, calculating the higher order corrections to the event weight
is vastly improved as the dimensionality of the phase space integration at NLO is
only three and independent of the jet multiplicity. 
This makes the integration tractable and straightforward to optimize
with a large variety of numerical integration options available.
Each colored line in the event has its own independent 3-dimensional integration.
The cancellations between virtual and soft/collinear can be fine-tuned per Born
event. One drawback of the method is that the subtraction
method has to match this approach. 
One has to ensure the subtraction term
projects back to the Born event and not to a variety of different Born events.
The more physically motivated slicing methods will not be affected by this issue since these methods
always project back to the originating Born event~\cite{Eynck:2001en,Harris:2001sx}.

In general the observable can be smeared by the \brem\!\!\!, as is shown in
Fig.~\ref{initfixed}. However this smearing is per construction infrared
unstable as it is generated by soft/collinear radiation pushing the
observable away from its Born value. 
Using the Born event momenta to define the observable will guarantee
an infrared insensitive observable and hence a stable prediction.
Predictions like the jet mass are best made using shower Monte Carlo's which can be easily matched
to the fixed LO event to fill back in the opaque jet with hadrons and
add the initial state hadronic radiation, providing a fully exclusive final state.
This will set up a natural separation between the fixed order perturbative part and
the shower part of the prediction, with higher order corrections providing a $K$-factor
to the showered LO event.

One can go a step further and produce an unweighted LO event sample, 
shower/hadronize the event and apply detector corrections. Next we can generate
for each of these LO events a $K$-factor using the FBPS generator
(see also ref.~\cite{Martini:2018imv}).
Assuming we have perturbative convergence, the $K$-factor should be of order one. This
will re-weight the showered and detector corrected Born event with a weight around one.
Applying radiative corrections in this manner is straightforward and time efficient.
The experimental jet algorithm does not need to be changed as long as the observable
is unaltered by the \brem radiation\!\!, i.e. the most exclusive observable is given by
$d\sigma/(d Q\{d\vec p_T^{i}d\eta_i\}_{i=1}^n)$ using the generator of Eq.~\ref{FBPScross}.
One can go further and change the clustering phase of the jet algorithm itself to obtain massless
jets and cluster the initial state radiation back into the final state objects. Then
the exact Born event is reconstructed and any observable using the Born event momenta
can be used to define observables.

The next phase of the project is to apply the FBPS generator to the NLO calculations of
$PP\rightarrow V+1$ jet, $PP\rightarrow 2$ jets and $PP\rightarrow V+2$ jets.
Each of these processes will add more complexity to the approach. After these processes
have been successfully computed in the FBPS approach, we can begin investigating and constructing NNLO processes such as
$PP\rightarrow V+1$ jet.

\acknowledgments 

T.~F. acknowledges the kind support of the Fermi Theory Group during several visits in 2017 and 2018.  
W.~G. is supported by the DOE contract DE-AC02-07CH11359. 

\bibliographystyle{JHEP}
\bibliography{mcfm_fbps_v1}
\end{document}